\documentclass{SciPost}
\pdfoutput=1

\hypersetup{
    colorlinks,
    linkcolor={red!50!black},
    citecolor={blue!50!black},
    urlcolor={blue!80!black}
}

\usepackage[bitstream-charter]{mathdesign}
\DeclareSymbolFont{usualmathcal}{OMS}{cmsy}{m}{n}
\DeclareSymbolFontAlphabet{\mathcal}{usualmathcal}

\fancypagestyle{SPstyle}{
\fancyhf{}
\lhead{\colorbox{scipostblue}{\bf \color{white} ~SciPost Physics Community Reports }}
\rhead{{\bf \color{scipostdeepblue} ~Submission }}

\fancyfoot[C]{\textbf{\thepage}}
}

\usepackage[utf8]{inputenc} % allow utf-8 input
\usepackage[T1]{fontenc}    % use 8-bit T1 fonts
\usepackage{hyperref}       % hyperlinks
\usepackage{url}            % simple URL typesetting
\usepackage{booktabs}       % professional-quality tables
\usepackage{nicefrac}       % compact symbols for 1/2, etc.
\usepackage{microtype}      % microtypography
\usepackage{xcolor}         % colors
\usepackage{tikz}
\usepackage{tcolorbox}
\usepackage{mdframed}

\usetikzlibrary{arrows.meta,fit,backgrounds,positioning}

\begin{document}

\pagestyle{SPstyle}

\begin{center}{\Large \textbf{\color{scipostdeepblue}{
Unknown Unknowns: Model Misspecification in Machine Learning for Physics
}}}\end{center}

\begin{center}\textbf{
Juan Cruz-Martinez\textsuperscript{1$\star$},
Carolina Cuesta-Lazaro\textsuperscript{2,3$\dagger$},
Alexander Held\textsuperscript{4$\ddagger$}, and
Michael Kagan\textsuperscript{5$\star\star$}
}\end{center}

\begin{center}
{\bf 1} Departamento de Física Atómica, Molecular y Nuclear, Universidad de Sevilla, E-41080 Sevilla, Spain
\\
{\bf 2} Center for Computational Astrophysics, Flatiron Institute, 162 Fifth Avenue, New York, NY 10010, USA
\\
{\bf 3} Institute for Advanced Study, 1 Einstein Drive, Princeton, NJ 08540, USA
\\
{\bf 4} University of Wisconsin–Madison, 1205 University Ave, Madison, WI 53706, USA
\\
{\bf 5} SLAC National Accelerator Laboratory, 2575 Sand Hill Rd, Menlo Park, CA 94025, USA
\\[\baselineskip]
$\star$ \href{mailto:jcruz@us.es}{\small jcruz@us.es}\,,\quad
$\dagger$ \href{mailto:ccuesta-lazaro@flatironinstitute.org}{\small ccuesta-lazaro@flatironinstitute.org}\,,\\
$\ddagger$ \href{mailto:alexander.held@wisc.edu}{\small alexander.held@wisc.edu}\,,\quad
$\star\star$ \href{mailto:makagan@slac.stanford.edu}{\small makagan@slac.stanford.edu}
\end{center}

\section*{\color{scipostdeepblue}{Abstract}}
\textbf{\boldmath{%
Machine learning is now a central tool for solving inverse problems in particle physics and astronomy.
Models are trained on simulation and deployed on real data, raising the question not just of whether they fit, but of whether they are wrong in ways we did not anticipate: the \textit{unknown unknowns}.
This challenge of \textit{model misspecification} is not unique to machine learning.
In physics, misspecification is sometimes exactly what we want to find: new discoveries appear as failures of existing models.
At other times, we want such effects absorbed into the analysis without biasing the measurement. A robust analysis is one that \textit{absorbs the misspecifications we are not interested in, while preserving sensitivity to the ones we are.}
Machine learning can both amplify misspecification and provide new tools to address it.
We discuss the challenges of model misspecification, diagnostics for detecting it, and strategies for mitigation.
No single diagnostic can confirm that a model is correctly specified: detection and mitigation are two halves of an iterative loop, in which a battery of complementary diagnostics is applied, the model is updated, and the process repeated.
Robustness against unknown unknowns is ultimately less about any single technique than about a disposition: a willingness to suspect one's own model, and to design analyses that can survive being wrong in ways one did not anticipate.
}}

\vspace{\baselineskip}

% %%%%%%%%%% BLOCK: Copyright information
% % This block will be filled during the proof stage, and finilized just before publication.
% % It exists here only as a placeholder, and should not be modified by authors.
% \noindent\textcolor{white!90!black}{%
% \fbox{\parbox{0.975\linewidth}{%
% \textcolor{white!40!black}{\begin{tabular}{lr}%
%   \begin{minipage}{0.6\textwidth}%
%     {\small Copyright attribution to authors. \newline
%     This work is a submission to SciPost Phys. Comm. Rep. \newline
%     License information to appear upon publication. \newline
%     Publication information to appear upon publication.}
%   \end{minipage} & \begin{minipage}{0.4\textwidth}
%     {\small Received Date \newline Accepted Date \newline Published Date}%
%   \end{minipage}
% \end{tabular}}
% }}
% }
%%%%%%%%%% BLOCK: Copyright information

%%%%%%%%%% LINENO
% For convenience during refereeing we turn on line numbers:
% \linenumbers
%%%%%%%%%% LINENO

\vspace{10pt}
\noindent\rule{\textwidth}{1pt}
\tableofcontents
\noindent\rule{\textwidth}{1pt}
\vspace{10pt}

%%%%%%%%%%%%%%%%%%%%%%
\section{Introduction}
%%%%%%%%%%%%%%%%%%%%%%

In 1998, two teams that had set out to measure how quickly
the expansion of the Universe was slowing down instead
discovered it was speeding up~\cite{Riess_1998,Perlmutter_1999}.
Their model of the cosmos, one without a cosmological
constant, was misspecified, and that misspecification turned
out to be one of the most important discoveries in modern
physics. Before dark
energy was accepted, both teams and the broader community
scrutinized every conceivable systematic
effect (dust, calibration, selection bias...) because
distinguishing genuine new physical phenomena from an
uncontrolled uncertainty is usually the hardest part.

This story is not an exception, it \textit{is} the
scientific method. Physics progresses by building models,
finding where they fail, understanding why, and iterating.
Most major discoveries began as a
misspecification: data that could not be explained by the
standard model used in the respective discipline.
In this sense, the central challenge of
physics has always been the one we address in this article:
how to detect model misspecification, how to determine
whether it matters, and how to respond to it.

The challenge is worsened by the nature of the measurements
themselves. In astrophysics and cosmology, we cannot design
controlled experiments; we observe a single Universe,
often through instruments whose limitations are difficult
to fully characterize. Simulations have become our
laboratories, allowing us to test hypotheses, calibrate
methods, and propagate uncertainties. But simulations are
themselves models, and any misspecification they carry
propagates silently into the analysis.

Machine learning (ML) has also become an integral part of this
pipeline~\cite{Carleo:2019ptp}. Across particle physics, astrophysics, and
cosmology, ML models now serve as fast emulators of
expensive simulations, as components of the inference
machinery, and as empirical models of physical processes
too complex for first-principles treatment. This
prevalence is driven by necessity: the data volumes,
dimensionalities, and complexities of modern experiments
increasingly exceed what traditional methods can handle.
But ML introduces its own risks. Models trained on
simulations inherit their misspecifications. Models that
learn correlations without causal structure can amplify
biases in out-of-distribution regimes. At the same time,
ML offers new opportunities: its ability to operate in
high-dimensional spaces makes it a powerful tool for
\textit{detecting} misspecification that would be
invisible to lower-dimensional diagnostics.
Flexibility in how ML models are defined
and trained additionally provides new handles for \textit{mitigating} it.

This article contributes to VERaiPHY (Validation \&
Evaluation for Robust AI in PHYsics), a PHYSTAT review
series establishing verification and validation standards
for machine learning across particle physics, astrophysics,
and cosmology. Elsewhere in this series, the application
of machine-learning techniques to physics, their
interpretation, and the quantification of uncertainties
have been discussed. Here we focus on robustness: what
happens when our models are wrong in ways we did not
anticipate? We argue that this challenge is not a
consequence of machine learning but is inherent to
physical modeling---and that the strategies physicists
have developed over decades to identify and mitigate
mismodeling apply equally to ML-based analyses.

We also argue that adopting a common language matters.
What physicists call ``systematics,'' statisticians call
``model misspecification,'' and ML practitioners call
``distribution shift.'' These are overlapping descriptions
of the same underlying problem, and using a shared
vocabulary, as we attempt throughout this
article, makes it easier to transfer insights across
communities and disciplines.

\begin{tcolorbox}[
  title=Key messages,
  colback=gray!5!white,
  colframe=gray!60!black,
  fonttitle=\bfseries
]
\begin{enumerate}
    \item Model misspecification is inherent to physics,
    not a consequence of machine learning.
    \item No diagnostic can confirm a model is correctly
    specified. We can only test for distributional
    adequacy, not mechanistic correctness.
    \item Machine learning amplifies both the risk
    and the detectability of misspecification, and provides new mitigation tools.
    \item Mitigation is iterative: detect, address,
    re-validate.
\end{enumerate}
\end{tcolorbox}

%%%%%%%%%%%%%%%%%%%%%%%%%%%%%%%%%%%%%%%%%%%%%%%%%
\subsection{What do we mean by misspecification?}
\label{sec:misspec-meaning}
%%%%%%%%%%%%%%%%%%%%%%%%%%%%%%%%%%%%%%%%%%%%%%%%%

Models in physics span a spectrum from \textit{mechanistic}, derived from first-principles physical laws, to \textit{empirical}, fit to data without mechanistic grounding. Empirical models---including scaling relations, phenomenological fits, and ML-based approximations---are common in physics, filling gaps where first-principles predictions are intractable or not well understood.

\begin{tcolorbox}[
  title=Definition: Model Misspecification,
  colback=blue!5!white,
  colframe=blue!40!black,
  fonttitle=\bfseries
]
We say a model $P_\theta$ is \textit{misspecified} for a
data-generating process $P^\star$ if $P^\star \notin \{ P_\theta:
\theta \in \Theta \}$, where $\theta$ are the parameters of the
model and $\Theta$ is the space of all parameters the model can
accommodate. Here $P^\star$ refers to the full data-generating
process---including its latent variables and causal
structure---not merely the distribution over observables. A model
is well-specified only when it contains the right causal structure.
\end{tcolorbox}

This is a deliberately strong definition, because in practice we can only probe a weaker condition: whether the model reproduces the observed data. The diagnostics we discuss in Section~\ref{sec:detecting} operate at this level. A model can pass every available goodness-of-fit test while remaining mechanistically wrong---producing the right distributions through the wrong process. Such a model is still misspecified in our sense, but its misspecification is invisible to observational tests. This gap between what we \textit{define} as misspecification and what we can \textit{detect} is a central tension throughout this article.

In physics, the path from theory to data typically involves a
chain of models: a physical theory governing the true state of
nature, an observation model mapping that state to what our
instruments can measure, and numerical implementations of both.
Misspecification can enter at any link in this chain, and when
a learned model is added for inference, it introduces yet another.

To make this concrete, it is useful to categorize the ways in
which each link can fail:

\begin{itemize}
    \item \textit{Structural}: the physical model itself is
    incomplete: missing degrees of freedom, wrong functional relationships, or incorrect causal structure in the underlying physical model. These are often difficult to diagnose as the model may appear to fit the data while encoding the wrong physics;
    \item \textit{Numerical}: approximations introduced when closed-form solutions are unavailable (discretization, finite resolution, sub-grid prescriptions). This connects to the above, with the subtlety that these are introduced due to known shortcomings and in a perfect world are known unknowns;
    \item \textit{Distributional}: incorrect specification of
    the statistical model (e.g., assuming Gaussian noise when errors are heavy-tailed, or misspecified priors).
    These should be distinguished from observational
    effects: the error is in the assumed form of the
    distribution, not in the instrument producing the data;
    \item \textit{Observational}: instrumental effects and selection functions that are difficult to characterize. To this category belong the systematic uncertainties introduced by the inherent limitations of the instruments, selection bias stemming from selection effects on a survey, etc. Some paradigmatic examples of mitigation efforts are noise modeling for the detection of gravitational waves~\cite{LIGOScientific:2019hgc} or the strategies to reduce the effect of pileup in the extremely crowded LHC collisions~\cite{CMS:2020ebo}.
\end{itemize}

\begin{tcolorbox}[
  title=Known unknowns vs unknown unknowns,
  colback=blue!5!white,
  colframe=blue!40!black,
  fonttitle=\bfseries
]
\textit{Known unknowns} are uncertainties whose existence we
recognize and can characterize: detector resolution,
perturbative truncation, finite sample size. They enter the
uncertainty budget and, when properly modeled, do not bias
the result.

\textit{Unknown unknowns} are forms of model inadequacy we
may not know exist. They cannot be captured by standard
uncertainty propagation and may introduce biases invisible
to conventional diagnostics.

The boundary between the two is not fixed: a known unknown
becomes effectively unknown when our characterization of
its impact is itself wrong.
Some mitigation strategies (see Section~\ref{sec:mitigate})
aim to turn unknown unknowns into known unknowns, e.g., by providing quantifiable uncertainties.
\end{tcolorbox}

In standard machine learning, model misspecification is typically a failure mode to be corrected: we train on data drawn from some distribution and hope our model generalizes well within a similar distribution. In physics the situation is more nuanced, because misspecification plays two very different roles depending on where in the modeling chain it sits and what we are trying to learn.

On one hand, we often wish to stress-test our models and find gaps in our understanding of nature; this is precisely what discovering new physics means. In other words, \textit{often our goal is to detect misspecification, not empirically correct it or hide it in any way.} Section~\ref{sec:detecting} focuses on diagnostics and tests that let us identify when a model fails.

On the other hand, much of the modeling chain describes effects we believe we understand in principle but cannot compute or simulate exactly: detector response, non-perturbative quantum chromodynamics, foreground contamination, finite-resolution simulations. These are closer in spirit to the out-of-distribution problem in machine learning. We do not expect insight from their misspecification, only contamination of the measurement we actually care about. Here the goal is mitigation: \textit{making our analysis robust to mismodeling we have no ambition to resolve}. Section~\ref{sec:mitigate} focuses on mitigation, strategies for ensuring that misspecification we cannot or do not need to fully understand does not bias our conclusions.

The two are complementary rather than opposed: a robust analysis is one that \textit{absorbs the misspecifications we are not interested in, while preserving sensitivity to the ones we are.}

%%%%%%%%%%%%%%%%%%%%%%%%%%%%%%%%
\subsection{Discovery or fluke?}
%%%%%%%%%%%%%%%%%%%%%%%%%%%%%%%%

The first challenge is assessing whether an anomaly is statistically significant or merely a random fluctuation. This problem has historically been addressed by adopting stringent significance thresholds. In particle physics, the convention is to require a $5\sigma$ significance before claiming discovery, corresponding to a p-value of approximately $3\times10^{-7}$. While this threshold may seem arbitrary, it has become a consensus after numerous $3\sigma$ and even $4\sigma$ anomalies have subsequently disappeared with more data. Notable examples include the 2003 pentaquark claims at LEPS, which reached $4.6\sigma$~\cite{LEPS:2003wug} before failing to replicate, and the 2015 diphoton excess at the LHC, which reached a local significance of $3.9\sigma$ before disappearing~\cite{ATLAS:2016gzy}. Ref.~\cite{Lyons:2013yja} provides a systematic discussion of the history of the $5\sigma$ criterion while arguing that the appropriate threshold should depend on the prior plausibility of the claim---with more extraordinary claims (such as physics beyond the Standard Model) potentially requiring even higher significance.

Yet finding a statistically significant anomaly still leaves the question of its origin open. The remaining challenge is to distinguish whether the misspecification is sourced by new physics or unmodeled systematic effects. An example of the latter is the 2014 BICEP2 announcement of primordial gravitational waves, later attributed to galactic dust contamination~\cite{Planck:2014dmk}. This ambiguity is particularly accentuated when systematic uncertainties may be underestimated. In such cases, the true significance of an anomaly is lower than the quoted value suggests, and what appears to be a robust detection may still be consistent with instrumental or modeling artifacts.

Similarly, the 2011 faster-than-light neutrino anomaly observed by the OPERA experiment was later attributed to faulty equipment affecting the timing system~\cite{opera_neutrinos_2012}.
The result came as a surprise and prompted additional checks and the desire for independent verification via the ICARUS experiment.
The scientific process continues after such an anomalous result, with the aim of achieving a deeper understanding of it.

The role of a physicist is then to come up with a new hypothesis that could explain an observed anomaly. This process of hypothesis generation might be termed ``interpretability'' in the machine learning sense: finding a mechanistic explanation for why the model fails where it does. There is emerging interest in whether large language models might assist in this hypothesis generation process for a fully automated discovery pipeline, though this is still very speculative.

It is important to distinguish our definition of model misspecification from outlier detection, a term sometimes used interchangeably with anomaly detection in the machine learning literature. Outlier detection typically asks whether an individual data point is rare under the model. But rarity alone is not evidence of misspecification: a rare event that occurs at its predicted rate is a successful prediction, not a failure. For example, gravitational wave signals from binary black hole mergers are rare, but they are detected at rates consistent with theoretical predictions. Misspecification, by contrast, occurs when events take place at the wrong rate, with unexpected properties, or in contexts where the model predicts they should not exist at all. In this sense, detecting misspecification is closer to what some authors mean by anomaly detection: identifying not rare events, but events that violate the model's expectations.

%%%%%%%%%%%%%%%%%%%%%%%%%%%%%%%%%%%%%%%%%%%%%%%%%%%%%
\subsection{Does machine learning make things worse?}
%%%%%%%%%%%%%%%%%%%%%%%%%%%%%%%%%%%%%%%%%%%%%%%%%%%%%

As we anticipated, our definition of model misspecification is not specific or restricted to machine learning applications, but generally holds in physics problems.
The addition of machine learning techniques to physics analyses can exacerbate model misspecification problems by enabling more powerful inference from complex data, while also opening avenues for more powerful new physics detection.

As ML models replace components of the physics analysis pipeline, they can learn predictive correlations without capturing the underlying causal structure, making misspecification worse in out-of-distribution regimes. This is compounded by approximation errors introduced through training data selection effects and architectural/optimization biases.

Moreover, detecting misspecification is only the first step. The scientific goal is to connect this detection to a physical hypothesis. In high-dimensional regimes where ML excels, understanding what is misspecified and why becomes substantially harder. This is not unique to ML: diagnosing misspecification in complex data has always been difficult, but ML's ability to leverage full field-level information means we increasingly operate in regimes where lower-dimensional intuition and visualization fail.

%%%%%%%%%%%%%%%%%%%%%%%%%%%%%%%%%%%%%%%%%%%%%%%%%%%%%%%%%%%%%%%%%%%%
\subsection{The brittle frontier between known and unknown unknowns}
%%%%%%%%%%%%%%%%%%%%%%%%%%%%%%%%%%%%%%%%%%%%%%%%%%%%%%%%%%%%%%%%%%%%

\begin{flushright}
{\it All models are wrong, but some are useful.} --- Attributed to George Box
\end{flushright}

Every physics analysis involves deliberate approximations. An
exact calculation may be intractable, so we truncate a
perturbative series. The full dataset may be too large or too
expensive to process, so we train on a subset. Our instruments
have imperfect calibration. In each case
we are knowingly introducing misspecification into the modeling pipeline; these are our known unknowns.

Known unknowns are, by themselves, not dangerous. As long as
their impact can be accurately characterized and propagated
through the analysis, they simply enlarge the uncertainty on
the final result without biasing it. A well-understood
detector resolution, for instance, does not prevent discovery;
it merely determines the precision with which we can make
a measurement.

The danger arises when our \textit{characterization} of a
known unknown is itself wrong. Consider the following examples:

At hadron colliders, predictions depend on the internal
structure of the proton, which is extracted from fits to
experimental data. The uncertainties on this structure are
themselves derived under specific assumptions, a choice of
functional form, a selection of which experiments to
include, a criterion for what counts as a good fit. If any
of these choices is inadequate, the resulting uncertainty
may not cover the true range of possibilities.

In cosmological surveys, the distances to galaxies are
often estimated from their observed colors rather than
measured directly. These estimates are calibrated against
smaller samples where precise distances are available. But
if the calibration sample is not representative of the full
survey, for instance, if it over-represents bright,
nearby galaxies, the calibration itself introduces a
systematic bias that may go unnoticed.

In each case the pattern is the same: the approximation is
known, but our confidence in controlling it is
misplaced.
The effect itself remains a known unknown, but the error in our characterization of it can constitute an unknown unknown:
a flaw whose existence we may be unaware of, which
may generate spurious anomalies, or worse, mask genuine ones.
Understanding how robust our analyses are in the presence of these effects is therefore essential and one of the main challenges in precision physics.

A paradigmatic example of the lack of a clear-cut boundary between fully controlled approximations and a misspecified model is the strong force in hadron colliders:
with our current knowledge of quantum chromodynamics (the theory which governs the strong interactions) we can only compute physical observables by performing a perturbative expansion on the coupling parameter of the theory, $\alpha_{s}$.
By computing, order by order, the terms in this series, we can control the theoretical accuracy of an observable.
However, the coupling parameter is not a constant, and its value changes with the energy of the collision.
At high energies, the coupling decreases (and the perturbation series is reliable) while at low energies the perturbation series breaks down and more terms (and non-perturbative effects) are necessary in order to obtain a reliable prediction.

Therefore any prediction calculated within this framework will have two sources of \textit{known} unknowns.
First, the missing contributions due to the truncation of the series: higher-order terms which we don't know how to compute.
We can, at best, estimate their effect or reduce their impact by focusing on highly energetic collisions.
Second, the low energy effects which cannot be computed via perturbation theory but that nonetheless might have a big impact on the calculation.
For the purposes of this discussion, let us focus on the latter.
In order to conduct any phenomenological study, we need a method to model these effects\footnote{While the concrete effects are not relevant to the discussion, for the purposes of grounding the explanation in physical effects one can think of the internal dynamics of the proton before the collision, the interaction of particles within the beam or the non-longitudinal components of the proton's momentum.}, a way to connect the model with the calculation and, crucially, a way to estimate the reliability of the model.
Note that while in this case the unknown is known, estimating its effect might not be.
Our predictions might be perfect (under our well controlled assumptions) but we might be over- (or under-)confident in them.
Continuing with the previous example, we might be over- (or under-)estimating the effect of the perturbative series and consider new physics an effect of the truncation of the series or otherwise claim discovery when what we are really seeing is an unmodeled higher-order effect.

With the advent of deep learning and the computational power that has made big data accessible, we have gained new ways of modeling these systems even when we lack enough theoretical insight as we can now treat these models as parametric black boxes that we can tweak (train) until they can explain enough physical phenomena to convince us they can generalize to new phenomena that the training algorithm has never seen.
While this is very powerful for scientific discovery, at the same time it exacerbates the ``unknownness'' of the unknowns: as we move further away from established theoretical insights it becomes increasingly difficult to validate our methods and estimate our own lack of knowledge.

%%%%%%%%%%%%%%%%%%%%%%%%%%%%%%%%%%%%%%%%%
\section{Detecting model misspecification}
\label{sec:detecting}
%%%%%%%%%%%%%%%%%%%%%%%%%%%%%%%%%%%%%%%%%

Given a model and data, how can we tell that something is wrong? This is the central question of this section, and
the honest answer is: not easily.
We emphasize from the outset that \textit{no single diagnostic is sufficient}.
Misspecification can
show up as a poor fit, as inconsistencies between
independent datasets, or as a failure under controlled
conditions---but it can just as well hide, masked by
degeneracies, limited statistical power, or the sheer
dimensionality of the problem.
The quadrant in Figure~\ref{fig:ood_quadrant} illustrates this: a model may pass goodness-of-fit tests globally while being misspecified in ways that degeneracies render undetectable (upper left), or a detected discrepancy may not significantly affect the physical quantity of interest (lower right).
No single diagnostic can cover all of these scenarios. We therefore advocate for a battery of complementary tests, each sensitive to different failure modes, applied together rather than in isolation.

A further complication is \textit{attribution}: when a
diagnostic flags a problem, it rarely tells us
\textit{where} in the modeling chain the problem
originates, whether the source is fundamental physics, the observation model, or the machine learning component of the analysis pipeline if present.
These are typically conflated in practice: all we observe is that the model, taken as a whole, does not adequately describe the data or generalize as expected.
The diagnostic power comes not from any single test, but from \textit{how intelligently the tests are designed}.
By choosing what to compare, which data subsets, which pseudodata variations, which summary statistics, we can progressively narrow the space of possible explanations and gain confidence about which modeling assumptions are driving the discrepancy.
In cases where a concrete hypothesis for the misspecification exists, such as a new physics model, we can devise tests that lead to a potential discovery.

With this in mind, we organize the discussion around
three families of diagnostics. \textbf{Goodness-of-fit testing}
(Section~\ref{sec:gof}) asks whether the model describes
the observed data. \textbf{Data splitting}
(Section~\ref{sec:splitting}) asks whether it generalizes
across physically distinct regimes. \textbf{Closure testing}
(Section~\ref{sec:closure}) asks whether the methodology
works under controlled conditions. Each probes a
different projection of the misspecification problem, and
each can catch failures that the others miss.

% Per-quadrant fills for Figure 1
\definecolor{quadYellowFill}{HTML}{FBF4E3}
\definecolor{quadYellowStroke}{HTML}{A87E1A}
\definecolor{quadCoralFill}{HTML}{FBEDE3}
\definecolor{quadCoralStroke}{HTML}{B85B2A}
\definecolor{quadGreenFill}{HTML}{ECF3E5}
\definecolor{quadGreenStroke}{HTML}{4F7A2E}
\definecolor{quadBlueFill}{HTML}{EEF3F9}
\definecolor{quadBlueStroke}{HTML}{0D2E5C}

\begin{figure}[htbp]
    \centering
    \resizebox{1.00\textwidth}{!}{%
    \begin{tikzpicture}[
        >={Stealth[length=3mm,width=2.5mm]},
        every node/.style={font=\normalsize},
        cell/.style={
            draw, rounded corners=2pt, line width=0.6pt,
            minimum width=4.4cm, minimum height=2.4cm,
            inner sep=2pt
        },
        quadlabel/.style={text width=3.9cm, align=center, font=\normalsize\itshape}
    ]
    % Quadrant fills (each cell its own color)
    \node[cell, fill=quadYellowFill, draw=quadYellowStroke!70] at (-2.85, 1.55) {};
    \node[cell, fill=quadCoralFill, draw=quadCoralStroke!70] at (2.85, 1.55) {};
    \node[cell, fill=quadGreenFill, draw=quadGreenStroke!60] at (-2.85, -1.55) {};
    \node[cell, fill=quadBlueFill, draw=quadBlueStroke!40] at (2.85, -1.55) {};

    % Crossed axes with arrows on both ends — dashed
    \draw[<->, line width=1pt, dashed] (-5.9,0) -- (5.9,0);
    \draw[<->, line width=1pt, dashed] (0,-3.4) -- (0,3.4);

    % Axis end labels
    \node[anchor=east, font=\normalsize\bfseries] at (-5.95, 0) {Low OOD detectability};
    \node[anchor=west, font=\normalsize\bfseries] at (5.95, 0) {High OOD detectability};
    \node[anchor=south, font=\normalsize\bfseries] at (0, 3.45) {High Model Misspecification};
    \node[anchor=north, font=\normalsize\bfseries] at (0, -3.45) {Low Model Misspecification};

    % Quadrant labels (original wording)
    \node[quadlabel] at (-2.85, 1.55)
        {Degeneracies mask misspecification};

    \node[quadlabel] at (2.85, 1.55)
        {Detected differences significantly affect the physical process of interest};

    \node[quadlabel] at (-2.85, -1.55)
        {No model misspecification};

    \node[quadlabel] at (2.85, -1.55)
        {Detected differences don't significantly affect the physical process of interest};
    \end{tikzpicture}%
    }
    \caption{Interpretations of possible scenarios of out-of-distribution (OOD) detectability and
    potential biases on inferred parameters due to model misspecification.}
    \label{fig:ood_quadrant}
\end{figure}
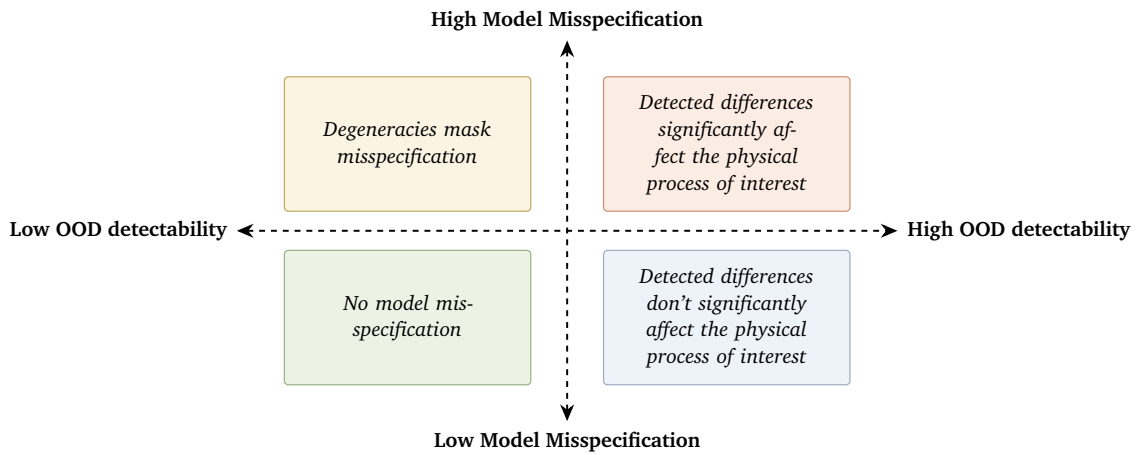

%%%%%%%%%%%%%%%%%%%%%%%%%%%%%%%%%%%%%%%%%%%%%%
\subsection{Goodness-of-fit}
\label{sec:gof}
%%%%%%%%%%%%%%%%%%%%%%%%%%%%%%%%%%%%%%%%%%%%%%

When confronting models with data, the first step towards the detection of model misspecification is to evaluate how well the model describes the observed data.
In the context of machine learning, this translates to evaluating whether the experimental data and data sampled from the model belong to the same underlying distribution. This is a weaker question than whether
the model is mechanistically correct, but it is the question our diagnostics can actually answer. As discussed in
Section~\ref{sec:misspec-meaning}, a model can pass every
distributional test while being wrong, reproducing the observed data through the wrong mechanism (``right for the wrong reasons'').

A range of metrics have been developed for this purpose,
and choosing among them involves trade-offs between
statistical power, computational cost, and
interpretability. Some of these metrics offer a numerical value which allows us to construct a hierarchy between models, while others provide only a binary answer. In Table~\ref{table:metrics} we present a quick overview of the metrics discussed below.

\begin{table}[ht]
    \centering
    \begin{tabular}{p{2.2cm} p{5.3cm} p{6cm}} \toprule
    Metric & Advantages & Caveats \\ \midrule
    C2ST & Classifier to distinguish data from model samples, simple and broadly applicable. & No insight about hierarchy of different models. \\
    OT & Cost of transforming between distributions. Offers a notion of geometric distance between models. & Very costly in high dimensions. \\
    SW & Scalable alternative to the metric above, takes one-dimensional projections. & By projecting we might be missing important correlations or features. \\
    MMD & Maps the sample to some kernel-defined feature space, bypassing the curse of dimensionality. & The choice of kernel acts as a bias. \\
    $\chi^{2}$ & Provides a straightforward metric for Gaussian-distributed histogrammed data. & Assumes statistically independent bins. \\
    correlated $\chi^{2}$ & Includes the covariance matrix in the calculation of the $\chi^{2}$, bypassing its main weakness. & A reliable estimate of the covariance matrix is often very challenging (and even the source of misspecification). \\
    \bottomrule
    \end{tabular}
    \caption{A bird's-eye comparison of common goodness-of-fit metrics.} \label{table:metrics}
\end{table}

When faced with two possible distributions, the
simplest approach is to ask a binary question: can
we tell samples from the two apart? The \textit{Classifier
Two-Sample Test} (C2ST)~\cite{osti_826696,lopezpaz2018revisitingclassifiertwosampletests} does exactly this by training a
classifier to distinguish model samples from data. If
the classifier performs significantly better than chance, the model is
detectably misspecified. While a powerful and simple way to flag a misspecified model, it offers no direct insight into the distance between the simulated and real samples. Any potentially small difference between the two datasets can be used to achieve very high classifier accuracy, and at the same time, its ability to distinguish between samples is sensitive to the training of the classifier.

When a notion of distance is required, one can use transport-based metrics. \textit{Optimal Transport} (OT) quantifies the minimum cost of transforming one distribution into another, yielding the Wasserstein or earth mover's distance. Unlike C2ST, this captures the geometry of the space: two distributions that are close in phase space will have a small transport cost even if they are statistically distinguishable. In high dimensions, computing the full Wasserstein distance becomes prohibitive, and the \textit{Sliced Wasserstein} (SW) distance~\cite{bonneel:hal-00881872} offers a scalable alternative by averaging over
one-dimensional projections.

Kernel-based approaches provide a simpler but less powerful alternative that scales well to high-dimensional comparisons. The \textit{Maximum Mean Discrepancy} (MMD)~\cite{JMLR:v13:gretton12a} computes the distance between the mean embeddings of two distributions within a Reproducing Kernel Hilbert Space, effectively bypassing the curse of dimensionality inherent to direct density estimation. This makes such kernel-based metrics particularly well-suited for the unbinned Monte Carlo simulations ubiquitous in particle physics~\cite{Kansal:2022spb}.

Finally, when data is naturally binned and Gaussian distributed, the familiar $\chi^2$ test provides a straightforward goodness-of-fit metric. However, the standard $\chi^2$ assumes statistically independent bins, which rarely holds in practice. The
\textit{correlated $\chi^2$} (or squared Mahalanobis distance) generalizes this by incorporating the full covariance matrix, capturing correlations between bins that a naive comparison would miss. This is often the natural starting point for analyses built around histogrammed observables, but it requires a reliable estimate of the covariance matrix, which itself can be a source of misspecification.

The right choice depends on the problem: its dimensionality, the available sample sizes, whether a binary answer or a quantitative distance is needed, and how much about the underlying distributions is known (stronger valid assumptions lead to more powerful tests).
But regardless of the metric chosen, a distributional comparison alone cannot be taken in isolation: it answers whether the model is detectably wrong, not whether it is wrong in ways that matter for the measurement at hand. The methods in the following sections complement these metrics by probing generalization (Section~\ref{sec:splitting}) and methodological robustness (Section~\ref{sec:closure}).

%%%%%%%%%%%%%%%%%%%%%%%%%
\subsection{Data splitting}
\label{sec:splitting}
%%%%%%%%%%%%%%%%%%%%%%%%%

Splitting the available data into training, validation, and test samples has become one of the most widespread and standard techniques when proposing, training, and validating machine learning models.
Its validity, however, extends beyond model evaluation.
To briefly summarize: when training a model we optimize a figure of merit or loss function.
During this optimization, the model is only allowed to see the ``training data,'' while the rest is held out.
To decide when to stop the optimization, the algorithm will check the figure of merit against the validation data. Once the validation metric no longer improves, the training is stopped to avoid overfitting.
Only then can the model be checked against the test data.
This 3-way split serves to reduce bias by preventing memorization from biasing performance estimates.

The same technique can also be adapted to estimate the robustness of a physical model or the validity of the physical assumptions applied to the model building.
In the scenario described above, the ideal split is one where all three samples are statistically representative of the full dataset, and a random split often serves this purpose well.
When evaluating a physical model, however, we will want to use theoretical insights also to guide how the data should be partitioned since we are interested in how the model will generalize to completely new and unseen scenarios.
Is the model truly following an underlying physical law, or is it merely capturing correlations specific to the experiments we have access to?
To test this one might split the data by energy regimes, different experiments, etc., rather than randomly.

Note that this principle is already employed, budget permitting, in major science projects.
For example, ATLAS and CMS, two of the LHC experiments, perform sibling analyses independently, with unpublished results kept internal to each collaboration.
This allows each to act as independent validation for the other.
Similarly, models developed for older experiments are regularly verified against completely new datasets.

In cosmology, the most powerful version of this
approach is the comparison between early and
late-universe probes. The cosmic microwave background
(CMB) provides a snapshot of the Universe roughly
$380{,}000$ years after the Big Bang. A cosmological model
fit to the CMB can then be used to \textit{predict} what
late-universe observations (supernovae, galaxy
clustering, weak gravitational lensing) should look
like. When these predictions agree with the late-universe
data, it is a stringent confirmation that the model
generalizes across billions of years of cosmic evolution.
When they disagree, it is a powerful signal that
something may be misspecified. The most prominent
example is the Hubble tension: the expansion rate
inferred from the CMB under the standard cosmological
model disagrees with direct
late-universe measurements using supernovae and other
distance indicators~\cite{Verde:2023lmm}. Whether this reflects
new physics or an uncontrolled systematic remains one of
the most actively debated questions in the
field.

These examples illustrate a general strategy: by
splitting data along physically meaningful boundaries,
we turn the comparison into a stress test of the
model's assumptions rather than merely a check against
overfitting.

%%%%%%%%%%%%%%%%%%%%%%%%%%%%
\subsection{Closure testing}
\label{sec:closure}
%%%%%%%%%%%%%%%%%%%%%%%%%%%%

Another popular technique to ensure the robustness of an optimization algorithm or machine learning methodology is using experimental pseudodata generated under fully controlled conditions.
This data should be as similar as possible to the experimental data, but sampled from a perfectly known distribution so that we can analytically affirm there are no unknowns.

Since in this case we know the underlying law and control the parameters of the distribution, we can check whether our model or methodology is able to explain it within a defined accuracy threshold.
If it does, we say that the methodology ``closes'' and that the closure test has succeeded, and we are confident that our methodology can also reliably model an unknown underlying law.
If it does not, it means there are aspects we cannot model and shortcomings to take into account when applying it to real data.
This technique effectively combines theoretical insights (as we need to define what the truth is) with statistical analysis (we often want to recover the underlying truth within uncertainties).

However, this technique is only as valid as our assumptions about the underlying law (in the best-case scenario, it is either known or its limitations are under control).
For instance, it is a very powerful technique to validate that the modeling of the background in experimental settings is accurate\footnote{For the sake of the discussion, we assume the background theory to be known.} despite complex experimental effects. Closure is also commonly used to test cosmological analyses via simulated universes.

Nonetheless, closure falls short when we need to model effects for which the underlying physics is largely unknown. This is the case, for example, in searches for new physics effects or modeling of non-perturbative contributions.

It has recently been proposed, in different contexts~\cite{DelDebbio:2021whr,AnauMontel:2024flo,Barontini:2025lnl},
that simulation-based distortions of input data can be used to quantify instances of mismodeling.
In particular, Ref.~\cite{Barontini:2025lnl} focuses on the impact of inconsistent input data on the methodology itself, but we can also flip the argument and consider the same set of checks to evaluate different theory assumptions.
By generating multiple sets of pseudodata (each of them governed by different theoretical settings) we can stress-test our model to quantify how much of a deviation from a given underlying theory it will absorb.
We can then choose to tune the sensitivity of our model to distinguish between theories and learn which kind of mismodeling it would be able to absorb and which can be reliably modeled without fear of losing important information.

%%%%%%%%%%%%%%%%%%%%%%%%%%%%%%%%%%%%%%%%%%%%%%%%%%%%%%%%%
\section{Mitigating the impact of model misspecification}
\label{sec:mitigate}
%%%%%%%%%%%%%%%%%%%%%%%%%%%%%%%%%%%%%%%%%%%%%%%%%%%%%%%%%

Section~\ref{sec:detecting} focused on detection: surfacing misspecification and attributing it to its physical source. This section turns to the complementary task of making our inference robust: \textit{sensitive to the effects we care about, insensitive to those we do not}. A robust analysis, in this sense, absorbs the misspecifications we are not interested in while preserving sensitivity to those we are.

In practice, detection and mitigation are not separable steps but two halves of the scientific method: the diagnostics of Section~\ref{sec:detecting} reveal issues, mitigation strategies address them, and the updated model must then be re-validated with the same diagnostics. Importantly, not every detected discrepancy should be mitigated. When attribution points to a candidate physical explanation, the appropriate response is to test that hypothesis, not to absorb the effect into the model. \textit{Mitigation risks hiding the very signals we are trying to find.}

\begin{figure}[htbp]
    \centering
    \resizebox{\textwidth}{!}{%
    \begin{tikzpicture}[
        >={Stealth[length=2.5mm,width=2mm]},
        box/.style={
            draw=quadBlueStroke, fill=quadBlueFill,
            rounded corners=2pt, line width=0.5pt,
            minimum width=2.4cm, minimum height=1.2cm,
            align=center, font=\small
        },
        pubbox/.style={
            draw=quadGreenStroke!70, fill=quadGreenFill,
            rounded corners=2pt, line width=0.5pt,
            minimum width=1.4cm, minimum height=0.8cm,
            align=center, font=\small
        },
        flow/.style={->, line width=0.7pt, draw=quadBlueStroke},
        annot/.style={font=\footnotesize\itshape, color=black!70}
    ]

    \node[box] (val) at (0,0)
        {\textbf{\textcolor{quadBlueStroke}{Validate / detect}}\\
         \textbf{\textcolor{quadBlueStroke}{misspecification}}};
    \node[box] (mit) at (4.5,0)
        {\textbf{\textcolor{quadBlueStroke}{Identify mitigation}}\\
         \textbf{\textcolor{quadBlueStroke}{strategy}}};
    \node[box] (imp) at (9,0)
        {\textbf{\textcolor{quadBlueStroke}{Implement (improved)}}\\
         \textbf{\textcolor{quadBlueStroke}{model}}};

    \begin{scope}[on background layer]
    \node[
        draw=quadYellowStroke!60, fill=quadYellowFill,
        rounded corners=4pt, line width=0.5pt,
        fit=(val) (mit) (imp),
        inner sep=12pt,
        inner ysep=28pt
    ] (blind) {};
    \end{scope}

    \node[anchor=north west, font=\footnotesize\itshape, color=black!75, inner sep=4pt]
        at (blind.north west) {Blind analysis to mitigate bias};

    \draw[flow] (val) -- (mit);
    \node[annot] at (2.25, 0.85) {Issue found};

    \draw[flow] (mit) -- (imp);

    \draw[flow] (imp.south) .. controls +(0,-2) and +(0,-2) .. (val.south);
    \node[annot] at (4.5, -2.5) {repeat until no issues remain};

    \node[pubbox] (pub) at (13.2,0) {\textbf{\textcolor{quadGreenStroke}{Publish}}};
    \draw[flow] (blind.east) -- (pub.west);
    \node[annot] at (12, 0.85) {All good!};

    \end{tikzpicture}%
    }
    \caption{Mitigation of model misspecification is an iterative process. Once issues are found, a mitigation strategy must be identified to result in an updated statistical model. The procedure repeats until no more problems can be found.}
    \label{fig:misspec-loop}
\end{figure}
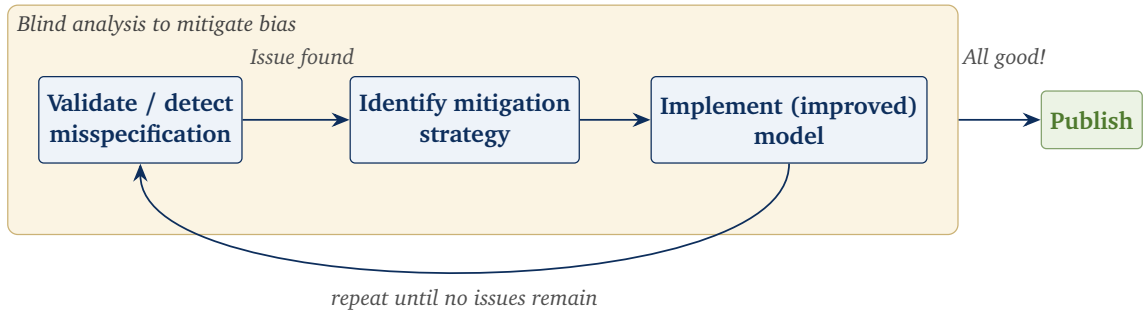

The strategies discussed in this section therefore apply to misspecifications we have judged to be systematic in origin, or which we cannot plausibly model from first principles. Figure~\ref{fig:misspec-loop} illustrates this iterative process.

No approach can fully protect against all unknown unknowns, and addressing whether the mitigation strategies for a given case are appropriate and sufficient generally requires careful contextual judgment from domain experts. The new model resulting from a mitigation step may still be affected by additional issues, so the loop continues until no more are found. Care should be taken when designing these tests and deciding how to act on their outcomes: testing for many possible aspects of misspecification will naturally yield some false positives, and whether to keep applying mitigation strategies requires consideration of both the number and types of tests done, as well as the motivation for the check and the plausibility of a related mismodeling~\cite{Barlow:2002yb}.

An important safeguard during this iterative process is the common practice of blind analysis~\cite{annurev.nucl.55.090704.151521}.
By hiding the final result of interest (such as the central value of a parameter being measured) until the loop stops and the statistical model is fully finalized, conscious and subconscious biases that may otherwise influence mitigation strategies can be avoided.
Methods for this can vary and include simply hiding the parameter during statistical inference, offsetting it by an unknown value, excluding the data most sensitive to the desired parameter, or replacing it with synthetic data instead to study the expected sensitivity without biasing decisions further.

Statistics literature refers to the underlying problem as \textit{post-selection inference}~\cite{Kuchibhotla:2022}.
Using the same data for selecting or modifying a statistical model and then subsequently drawing conclusions from it can lead to issues such as biases or miscalibrated coverage.
Blind analysis can be similar in spirit to the idea of sample splitting: by finalizing a model using only data insensitive to the parameter(s) of interest in a given measurement, bias can be mitigated.
However, a clean split into data used for model selection and data used for subsequent inference is not always performed, as frequently data from background-enriched regions used for model selection may still enter the final inference stage.

Perhaps the most intuitive way to approach a discussion of mitigation options is to start from a simulation-based perspective frequently encountered in physics~\cite{cranmer2015practicalstatisticslhc}.
In this case, a detailed model of nature is available, implemented with high-fidelity simulators.
Misspecification arises due to imperfections in those simulators.
As Section~\ref{sec:misspec-meaning} describes, a variety of misspecification types can exist and require mitigation, such as structural issues (e.g., using a leading-order approximation) or numerical issues (e.g., finite Monte Carlo sample size).
Many of the same mitigation ideas described in the following apply when the statistical model is specified without a simulator.
In exchange for avoiding potential simulator issues, this can complicate the \textit{attribution} process, which is the process of tracing an observed issue to its source to turn an unknown unknown into a known unknown (or ideally a known known).
Attribution can be simpler when identifying shortcomings in a mechanistic system compared to interpreting inadequacy of a functional form describing a distribution empirically.

%%%%%%%%%%%%%%%%%%%%%%%%%%%%%%%%%%
\subsection{Mitigation strategies}
%%%%%%%%%%%%%%%%%%%%%%%%%%%%%%%%%%

The preferable approach to tackling model misspecification is improving the original model directly at the source by computing higher perturbative orders, increasing the resolution of a simulation, or refining an instrument model.
Removing misspecification in this manner is often not possible, so the remainder of this section is focused on mitigation strategies to employ instead.
We distinguish four categories of such strategies.

\begin{tcolorbox}[
  title=Mitigation strategies,
  colback=blue!5!white,
  colframe=blue!40!black,
  fonttitle=\bfseries
]
\begin{itemize}
    \item \textbf{Covering with uncertainties}:
        Incorporating appropriate terms into the statistical model can help cover systematic errors due to model misspecification.
    \item \textbf{Calibration to correct the model}:
        With the use of reweighting and similar strategies, simulations can be adjusted to more accurately reflect nature.
    \item \textbf{Avoiding misspecified features}:
        If misspecification can be localized, we can construct robust models that are designed to be insensitive to it.
    \item \textbf{Empirical and data-driven modeling}:
        Empirical models and data-driven approaches can help sidestep the problem by avoiding model misspecification due to imperfect simulators.
\end{itemize}
\end{tcolorbox}

Their applicability depends on the category of misspecification described in Section~\ref{sec:misspec-meaning}.
As a rough guide, numerical and observational misspecification can often be understood well enough to be covered with uncertainties or corrected through calibration.
Structural and distributional issues on the other hand tend to call for avoiding misspecified features or working with empirical/data-driven models instead.

The mitigation strategies described are not unique to ML-based or -supported measurements.
The use of ML can both increase the impact of misspecification and help address it more effectively.
Consider the task of distinguishing between a hypothetical beyond the Standard Model process containing new physics and background contributions.
If simulator defects lead to differences between these processes, an ML model may learn to exploit them.
The resulting misspecification may be amplified when this happens across a high-dimensional phase space, compared to hand-crafted summary statistics using fewer features.
At the same time, ML-based reweighting strategies may help correct the simulation more effectively than hand-crafted low-dimensional reweighting approaches that cannot capture the relevant higher-dimensional phase space.

The four categories of mitigation strategies are not mutually exclusive and can be used jointly to address a given misspecification concern.
Finding the right approach in a given situation while iterating through the loop depicted in Figure~\ref{fig:misspec-loop} can additionally benefit from sensitivity analysis.
This helps quantify how strongly a result can depend on a specific modeling choice and thereby guide where effort should be focused.
Differentiable analysis pipelines~\cite{DeCastro:2018psv,Simpson:2022suz}, where the measurement is implemented as a differentiable program, are an emerging approach for gradient-based end-to-end sensitivity analysis.

The choice of inference paradigm can also play a role for misspecification.
Simulation-based inference (SBI)~\cite{doi:10.1073/pnas.1912789117} allows inference directly on a high-fidelity simulation chain, treating it as an implicit statistical model.
This can bypass summary statistics approximations used otherwise, but motivates SBI-specific diagnostics and mitigation strategies to deal with possible simulator misspecification issues~\cite{Cannon:2022,Wehenkel:2024,AnauMontel:2024flo,Pierre:2025ulp}.
These references include examples of field-specific applications, such as for gravitational waves or galaxy clustering.
The SBI review from the same series as this paper goes into more detail on the topic~\cite{Dax:2026qsm}.

No matter the approach chosen, we recommend documenting all assumptions that go into a statistical model clearly alongside the misspecification measures applied.
Future findings can always unveil sources of misspecification not considered at the time of publication, but being transparent about assumptions and the tests performed will help the reader interpret a publication even when viewed in light of future discoveries.

%%%%%%%%%%%%%%%%%%%%%%%%%%%%%%%%%%%%%%%%
\subsection{Covering with uncertainties}
\label{sec:cover-unc}
%%%%%%%%%%%%%%%%%%%%%%%%%%%%%%%%%%%%%%%%

When misspecification is characterizable and quantifiable, we can attempt to \textit{cover} it by incorporating appropriate degrees of freedom in our model.
This elevates a potential \textit{unknown unknown} to a \textit{known unknown}, usually described in an approximate fashion to avoid over-confident results.

Incorporating nuisance parameters into the statistical model is common practice to describe sources of uncertainty.
This can be done to describe known and well-understood effects, such as detector efficiencies obtained from calibration measurements or imperfect knowledge of fundamental constants of nature.
These examples are not strictly speaking misspecification issues: once incorporated into a statistical model, they parameterize \textit{known unknowns} which can be propagated through inference machinery and profiled or marginalized.
See~\cite{Dorigo:2020ldg} for a comprehensive review of ML-based approaches to nuisance parameters in high energy physics, and Ref.~\cite{Haussmann:2026gbi} for a review from this same series on uncertainties.
This includes inference-aware techniques that can recover sensitivity lost to parameterized nuisances but do not address potential misspecification in the parameterization itself.

Uncertainties are classified into three classes in~\cite{Sinervo:2003}.
Those statistical in nature, coming for example from in-house calibrations, are also sometimes called ``good.''
They are reducible with more data and parameterize a \textit{known unknown}.
The second class concerns sources of uncertainty where some assumptions need to be made about the measurement technique that cannot be verified, also referred to as ``bad.''
The third and last class of ``ugly'' systematics contains theory-related sources of uncertainty.
A statistical interpretation of these can be very difficult, for example when comparing two different approximate calculations of an expected outcome of a measurement.
Uncertainties introduced specifically to cover potential misspecification will typically be of the second or third kind, where exact details of the size or shape of an effect are uncertain or unknown.
We stress the challenge in designing a statistical model that can handle such sources of uncertainty while aiming for well-calibrated results.
Care must be taken in this process and overly inflated uncertainties can not only result in sensitivity losses, but also complicate the interpretation of the result~\cite{Barlow:2002yb}.

Common examples from particle physics which fall into the third class are factorization and renormalization scale variations, used as an approximate way to describe missing higher-order corrections in perturbative calculations.
Varying these scales by factors agreed upon by convention provides a way to turn this into a \textit{known unknown}.
The answer for how well this approximation works can in practice only be found when higher-order calculations become feasible in the future.
More sophisticated methods have recently been proposed~\cite{Bagnaschi:2014wea,Duhr:2021mfd,Tackmann:2024kci} for a more principled approach to the problem.

In some cases we may have access to multiple simulators which differ in their predictions, but we cannot necessarily exclude any of them as a wrong model of nature.
The difference between these models can then be taken into account~\cite{ATL-PHYS-PUB-2019-043} to define uncertainties.
A common approach at the LHC is to define a smooth interpolation between multiple discrete predictions from different simulators.
The interpretation of such an interpolation between discrete choices is difficult, underlining the challenge inherent to these ``ugly'' systematics: they are a pragmatic solution to an ill-defined problem.
Adding such uncertainties to a model needs to be done with the awareness of the assumptions being made, including also how different effects factorize.
The ultimate reliability of this method of covering mismodeling issues with uncertainties depends on how well the space of plausible mismodelings is parameterized.
Sources of mismodeling not anticipated may end up not being covered by any nuisance parameters, motivating the usage of complementary strategies.

%%%%%%%%%%%%%%%%%%%%%%%%%%%%%%%%%%%%%%%%%%%%%
\subsection{Calibration to correct the model}
\label{sec:calibration}
%%%%%%%%%%%%%%%%%%%%%%%%%%%%%%%%%%%%%%%%%%%%%

When we have a model for our data with known shortcomings, ad hoc corrections may be desirable to reduce the impact on our measurement.
Calibration measurements to characterize a detector apparatus are common and can subsequently be included with suitable nuisance parameters as described in Section~\ref{sec:cover-unc}.
For other sources of misspecification, an ad hoc correction via a calibration procedure may help.

In scenarios where the statistical model predicts a distribution based on simulated events, a reweighting procedure can be applied.
By scaling individual events, a distribution can be calibrated to match bin-by-bin with a reference distribution in the data.
This \textit{vertical} calibration can also be performed for more complex multivariate observables.
In higher dimensions, a classifier-based reweighting can provide per-event weights via the likelihood-ratio trick by training the classifier to distinguish between events in a source (e.g., a simulator) and target (e.g., observed data) distribution~\cite{cranmer2016approximatinglikelihoodratioscalibrated,Rogozhnikov_2016}.
Reweighting can only work when the simulated events have sufficient support, as they cannot populate regions of phase space where no events exist.

The second approach to calibration proceeds in a \textit{horizontal} fashion by adjusting the properties of individual events.
Normalizing flows~\cite{algren2023flowawaydifferencesconditional,Daumann_2024} and optimal transport~\cite{Pollard:2021fqv,Algren:2025zff} are approaches used to accomplish this.
Particularly in high-dimensional spaces, where lack of support can become an increasing concern, horizontal calibration can provide a more promising approach.
Ref.~\cite{ATLAS:2025rbr} shows a detailed practical example of an optimal transport approach applied to flavor tagging calibration in the ATLAS experiment.

Instead of calibrating events originating from running a chain of simulators, calibration can also be performed upstream and proceed as a \textit{tuning} of simulator parameters against control measurements.
The same idea can also be used to calibrate the physics model itself, for example to characterize hadronization models from data~\cite{Assi:2025avy}.

While calibration adjusts simulation to match data, unfolding operates in the opposite direction and corrects observed distributions for detector effects to infer particle-level distributions~\cite{Huetsch:2024quz}.
This relocates potential misspecification issues: an unfolded measurement can be compared with improved predictions after publication, but retains the detector model and a potential bias towards the simulation used.
Reducing this bias is an area of active development~\cite{Backes:2022sph,Butter:2025via}.

No matter the approach to calibration, the pursuit of this requires sufficient data to derive a correction alongside its associated uncertainty.
A low-dimensional correction that improves one observable may have adverse effects on the modeling of other observables.
At the same time, calibrating in a high-dimensional space can also risk absorbing misspecification of new physics effects that we may be interested in.
It is therefore important to pair calibration with checks that it performs as intended with control data.

%%%%%%%%%%%%%%%%%%%%%%%%%%%%%%%%%%%%%%%%%%%
\subsection{Avoiding misspecified features}
%%%%%%%%%%%%%%%%%%%%%%%%%%%%%%%%%%%%%%%%%%%

Instead of correcting for misspecification, an analysis can be designed to minimize dependence on information that may be poorly modeled.
This may be preferable when a calibration would be difficult, for example due to the lack of sufficient data.
Like the previous approaches, this requires us to suspect where misspecification may originate.
It cannot guard against unknown unknowns whose direction we cannot anticipate, but it can shrink the space of misspecifications that affect the result.
In general ML, many of the ideas captured here fall into the topic of \textit{domain adaptation}.

We may have reasons to assume that certain features of our analysis are not to be trusted: predicted low-energy behavior where perturbative approximations break down, specific poorly understood regions in our detector, data taken during a partial power outage.
The most straightforward approach is to omit certain features or observations.
In practice, this can mean not fitting a differential distribution for which we know our prediction is likely wrong, or not including the corresponding observable in the training of a discriminator.
It can also mean only selecting data taken while our detector and experimental setup were in a good state.

Simply excluding such features may not be enough.
A sufficiently expressive ML model, when provided with a set of observables characterizing a particle collision, may still be able to implicitly reconstruct features that we did not explicitly provide.
If such features are relevant to the task at hand---such as discrimination between different classes---but also affected by misspecification, a more explicit approach to suppress such dependence is desirable.

The adversarial ``learning to pivot'' approach~\cite{Louppe:2016ylz} trains both a classifier and an adversary simultaneously.
The adversary attempts to predict the value of a nuisance parameter, an unknown describing a family of plausible scenarios which we do not want to depend on, from the output of the classifier.
By introducing an adversarial penalty, we obtain a handle to steer between accuracy and robustness along the direction of the chosen nuisance parameter.
Another way to approach this is based on distance correlation, replacing the adversary with a penalty term that decorrelates the classifier output from the nuisance parameter~\cite{Kasieczka:2020yyl}.
While decorrelation approaches may be desirable for robustness, they can in some cases also result in lower sensitivity than classifiers where the dependence on the nuisance parameter is retained~\cite{Ghosh:2021roe}, for example by making the dependence explicit through a parameterized neural network~\cite{Baldi:2016fzo}.

When training data from several simulators with plausible settings is available, all of which are compatible with current observations, a simpler non-adversarial approach is to incorporate all of them in the training.
This will in general lead to a model that is more robust against these variations by forcing it to rely on features that are stable across them.
It may still be desirable to incorporate these different predictions into the statistical model with uncertainty terms as described in Section~\ref{sec:cover-unc}, but the impact of these variations on the analysis can be reduced with this approach.
This is closely related to the ideas of \textit{data augmentation} in general ML and \textit{domain randomization} in robotics.
Domain randomization proceeds by randomizing simulator parameters in robot training, leading to learned policies that better generalize to reality.
The same training-time machinery can also be used to encode invariance to a known symmetry, by augmenting the training data with corresponding transformations.
For example, a discrimination task in a rotationally symmetric experiment should not depend on the absolute orientation of the event, which can be enforced by applying random rotations to the training data.
Related techniques such as contrastive learning, pulling together events that differ only by chosen augmentations, can also be used to increase robustness against systematic variations~\cite{Wilkinson:2025nxv}.
More generally, self-supervised learning can produce representations insensitive to a chosen set of transformations, from physics-motivated augmentations~\cite{Dillon:2021gag} to re-simulation with varied simulator settings~\cite{Harris:2024sra}.
It also underlies foundation-model-style pretraining, for example via masked particle modeling~\cite{Golling:2024abg}.
The representation learning review from this report series~\cite{VERaiPHY_representation} provides more details.

Robustness can also be built directly into the architecture of ML models, which provides a stronger guarantee than augmentation.
A symmetry encoded in the architecture is respected exactly, rather than learned approximately from augmented samples.
This represents an injection of inductive bias, which may serve a dual role: improving the power of the model and the sensitivity of our analysis, while simultaneously preventing the model from exploiting misspecified features that violate known symmetries.
Physics-informed architectures can respect known symmetries by construction, such as permutation invariance over reconstructed jets in a particle collision~\cite{Fenton:2020woz}, Lorentz equivariance for four-momenta~\cite{pmlr-v119-bogatskiy20a,Gong:2022lye,Brehmer:2024yqw}, or combinations of multiple symmetries~\cite{Bogatskiy:2022czk}.
Many more examples are provided in the HEP ML Living Review~\cite{hepmllivingreview}
and the review on symmetries from this report series~\cite{VERaiPHY_symmetries}.

Beyond the model itself, the design of the target measurement can also be chosen to minimize sensitivity to misspecification.
A common idea experimentally is to measure ratios of observables affected by similar sources of uncertainty.
When numerator and denominator share a common misspecification, it can cancel to leading order.
This not only may reduce the impact of corresponding uncertainties attached to the model as described in Section~\ref{sec:cover-unc}, but has the potential to even mitigate the effect of unknown unknowns that may cancel in such a setup.
Many examples of this exist, such as the measurement of the $W$ and $Z$ boson production cross section ratio in association with a jet by the ATLAS Collaboration~\cite{ATLAS:2011kws}, specifically designed to maximize the cancellation of (known) experimental and theoretical uncertainties.
This approach is complementary to others described above: it does not require precise knowledge of the form of misspecification, only that it is shared between the processes being compared.

%%%%%%%%%%%%%%%%%%%%%%%%%%%%%%%%%%%%%%%%%%%%%%%%%%%%%%%%
\subsection{Empirical and data-driven modeling}
\label{sec:data-driven}
%%%%%%%%%%%%%%%%%%%%%%%%%%%%%%%%%%%%%%%%%%%%%%%%%%%%%%%%

One way to avoid issues with misspecification in simulation is to reduce or remove dependence on it, instead relying on observed data directly where feasible.
Such data-driven approaches shift the assumptions we make: instead of relying on a simulator to model nature accurately, we make assumptions about how we can extract distributions required for our measurement directly from observed data.
This approach works as long as those assumptions hold; misspecification issues can appear when that is not the case.
The decision to pursue this mitigation strategy therefore should take into account which assumptions may be more justified in a given case.

A common example is the search for a resonance on top of a smooth spectrum.
At the LHC, this includes searches for diphoton resonances on top of a smoothly falling background in a diphoton invariant mass distribution.
The 2015 diphoton excess at 750 GeV is a famous case of this where the ATLAS and CMS collaborations published results showcasing a local excess that hinted at a possible new particle, which subsequently disappeared when analyzing more data~\cite{ATLAS:2017ayi,CMS:2016kgr}.
By assuming that the background distribution falls within a suitably chosen family of distributions, an effective model can be determined from sideband measurements that exclude the region where a potential signal resonance is located.
These assumptions may be violated: contributions from signal or other processes that do not follow the same distribution may still be present in the sideband, or detector geometry may change across the spectrum in ways that distort it locally.
If that is the case, this data-driven background model can also be misspecified, causing the need for additional mitigation measures.
Hybrid approaches are possible: for example, when the differential distribution of a process is modeled well by simulation but its absolute rate is not, the rate can be measured from data in situ and the shape can be taken from simulation.

The choice of functional form to describe distributions can sometimes be motivated by theoretical physics considerations, while in other cases it proceeds ad hoc.
This is a place where ML-based approaches can help: non-parametric models provide a way to fit distributions by starting from different assumptions.
Gaussian processes can describe a broad space of functions given a choice of kernel and mean~\cite{Frate:2017mai}.
Normalizing flows can play a similar role for higher-dimensional distributions, making them suitable, for example, in the context of anomaly detection~\cite{Hallin:2021wme}.
The need for some assumptions to be made remains in the choice of flow architecture.

Ideas for the use of ML are not limited to describing smooth distributions of backgrounds.
The classification without labels (CWoLa) approach provides a framework for training a classifier on data to distinguish mixtures of classes~\cite{Metodiev:2017vrx}, thus avoiding the risk of learning features from imperfect simulators.

Data-driven techniques can extend beyond providing a prediction for a distribution in a narrow window from sidebands.
A very common approach is the ABCD method, dissecting signal and background processes into four disjoint regions by cutting on two observables.
Background contributions in the signal region can then be estimated from the remaining three background-rich regions.
Conceptually, ABCD is a reweighting similar to the calibration approach of Section~\ref{sec:calibration}, with the source distribution drawn from data rather than simulation.
ML can help here to identify suitable observables~\cite{Kasieczka:2020pil} to cut on.
Another common need is the description of fake and non-prompt leptons at the LHC.
As these processes are challenging to model with simulation, many data-driven techniques have been developed to avoid simulation-based misspecification issues~\cite{ATLAS:2022swp}.

A common feature across all these data-driven techniques is that they cannot eliminate the need to make potentially wrong assumptions that lead to misspecification, but they can relocate the assumptions.
The best approaches in a given situation are those where the assumptions are defensible on physics grounds, verifiable in control samples, and detection of other sources of misspecification as described in Section~\ref{sec:detecting} remains possible.

%%%%%%%%%%%%%%%%%%%%%%%%%%%%%%%%%%%%%%%%%%%%%%%%%%
\section{Towards automating the scientific method}
%%%%%%%%%%%%%%%%%%%%%%%%%%%%%%%%%%%%%%%%%%%%%%%%%%

A more speculative direction at the boundary of data-driven and mechanistic modeling is the use of large language models (LLMs) to assist, or eventually automate, parts of the scientific method itself. The vision is one in which the loop of Figure~\ref{fig:misspec-loop} (and indeed much of what we have discussed in this article) becomes machine-driven: hypothesis proposal, model construction, validation, and revision performed by an automated system rather than a human analyst. Several developments in the last two years suggest this is no longer purely aspirational. AlphaEvolve~\cite{novikov2025alphaevolvecodingagentscientific} pairs large language models with evolutionary search and has produced concrete results in mathematics and algorithmic design; more recently Bayesian formulations recast model discovery as posterior inference over a space of mechanistic simulators~\cite{wahl2026probabilisticframeworkllmbasedmodel}.

The two approaches differ in how they score candidate solutions. AlphaEvolve treats discovery as an evolutionary search: the LLM proposes code, an evaluator function returns a fitness score, and successful candidates are recombined and mutated to seed the next generation. It has excelled in domains where the evaluator is unambiguous, such as breaking a 56-year-old record set by Strassen's algorithm for multiplying $4\times4$ complex-valued matrices and improving bounds on the kissing number in eleven dimensions~\cite{novikov2025alphaevolvecodingagentscientific,georgiev2025mathematicalexplorationdiscoveryscale}. The Bayesian formulation of Ref.~\cite{wahl2026probabilisticframeworkllmbasedmodel} replaces fitness with weighting models: candidate simulators proposed by the LLM are sampled in proportion to how well they explain observed data, using a Sequential Monte Carlo procedure. The shift is conceptually meaningful for physics, where the evaluator is rarely as clean as a mathematical bound: making the uncertainty over the model itself an explicit output of the procedure is closer to how a physicist would actually want to think about model discovery in the presence of noise. Two issues, however, remain open.

The first is that there is no clear metric for what scientific discovery actually \textit{is}. Predictive accuracy on held-out data is the obvious choice, and it is what most current systems optimize for, but it is not what physicists mean by understanding a phenomenon. Lorentz transformations were originally introduced ad hoc to accommodate the results of electromagnetic experiments before Einstein recognized that they followed from simple principles, notably the invariance of the speed of light. The same data admits both descriptions, and both predict the same observations, but only one generalizes effortlessly to regimes its authors had not considered and opens up a new field for science.

The second, related issue is that current LLM-based discovery systems do not appear to have \textit{taste}. By taste we mean the physicist's intuition for which of several empirically adequate explanations is more likely to generalize, a sense for symmetry, parsimony, and conceptual coherence that goes beyond fit. This is not a mystical capability; much of it is captured by Occam's razor, by preference for explanations that connect to existing theoretical structure, and by skepticism toward laws that (over-)fit one regime through the wrong mechanism. But it is precisely the part of the scientific process least likely to be learned from text, and the part most likely to be missing when we ask an automated system to propose models.

Where these systems are most likely to help in the near term is in the brute-force exploration of misspecification scenarios that humans rarely run. Most of the strategies described in Section~\ref{sec:detecting}, generating alternative simulators, varying functional forms, stress-testing models against pseudodata under deliberately different theoretical assumptions, are limited in practice not by methodology but by human effort. An automated system can explore far more of this space than a single collaboration can, and it does not need taste to do so usefully: it needs only to generate candidates that are diverse and runnable. The harder question, and the one where the absence of taste matters most, is what to do with the results. The same systems that can propose a hundred candidate misspecifications can also propose a hundred plausible-looking explanations for any apparent anomaly, and the burden of distinguishing the genuinely interesting ones still falls on the human researcher. As with the data-driven approaches described in Section~\ref{sec:data-driven}, automation does not eliminate the need to make potentially wrong assumptions, it relocates them onto the LLM's prior over physically sensible models and onto the validity of the likelihood used to score them, neither of which is easy to inspect or constrain.

%%%%%%%%%%%%%%%%%
\section{Summary}
\label{sec:summary}
%%%%%%%%%%%%%%%%%

\begin{flushright}
{\it The first principle is that you must not fool
yourself---and you are the easiest person to fool.}
--- Richard Feynman
\end{flushright}

Misspecification plays two roles in physics, and conflating them is a mistake. Sometimes the failure of a model \textit{is} the discovery: new physics manifests as a model that no longer explains the data. Sometimes the failure is contamination to be absorbed, such as detector effects, non-perturbative corrections, or foreground emission, where we have no ambition to resolve the misspecification, only to prevent it from biasing the result. A robust analysis is one that preserves sensitivity to the former while absorbing the latter, and most of the methodological choices we have discussed come down to drawing that distinction well.

Moreover, misspecification is not a problem introduced by machine learning. Physics has always progressed by finding where models fail, and the strategies developed over decades to identify and mitigate mismodeling apply directly to ML-based analyses. What machine learning changes is the scale of the problem: high-dimensional inference can both amplify the impact of simulator defects and surface failures invisible to lower-dimensional diagnostics.

A point we have stressed repeatedly is that no diagnostic can confirm a model is correctly specified. We can only test for distributional adequacy, never mechanistic correctness, and a model can pass every available test while being right for the wrong reasons. The boundary between known and unknown unknowns is also less stable than it appears:
the moment our characterization of a known unknown's impact is itself wrong, that mischaracterization is an unknown unknown in its own right,
as the history of uncertainties on parton distribution functions, photometric-redshift calibrations, and analogous cases makes clear. We therefore advocate for a battery of complementary diagnostics rather than reliance on any single test, recognizing that detection rarely tells us where in the modeling chain a problem originates: attribution comes from the intelligent design of tests, not from the tests themselves.

Detection and mitigation are not separable steps but two halves of an iterative loop. Once a discrepancy is found, it must be attributed, addressed, and the updated model re-validated with the same diagnostics. Not every discrepancy, however, should be mitigated. When attribution points to a candidate physical explanation, the appropriate response is to test that hypothesis, not to absorb the effect into the model. Mitigation strategies (covering with uncertainties, calibrating the model, avoiding misspecified features, working from data directly) are tools for the misspecifications we judge to be systematic in origin, and they should be deployed with the awareness that they risk hiding the very signals we are trying to find. Data-driven techniques in particular do not eliminate assumptions; they relocate them, and the best choice in a given situation is the one whose assumptions are most defensible on physics grounds and verifiable in control samples.

Finally, robustness against unknown unknowns is not a property that can be fully guaranteed. Blind analysis remains an essential safeguard against the biases that creep in when the same data is used to select and to interpret a model. Robustness is ultimately less about any individual technique than about an attitude: a willingness to suspect one's own model, and to design analyses that can survive being wrong in ways one did not anticipate.

\paragraph{Acknowledgments} The authors thank G. Grosso and R. Winterhalder for their support and coordination of the VERaiPHY project. We also thank M. Dax and J. Spinner for their reviews, and J. Linhart for her comments on our paper. J.C.-M. acknowledges funding from the Ramón y Cajal program grant RYC2023-043794-I funded by MCIN/AEI/10.13039/501100011033 and by ESF+. A.H. is supported by the US National Science Foundation (NSF) cooperative agreements OAC-1836650 and PHY-2323298 (IRIS-HEP). M.K. is supported by the US Department of Energy (DOE) under Grant No. DE-AC02-76SF00515.

\bibliography{main}{}

\begin{thebibliography}{10}
\providecommand{\url}[1]{\texttt{#1}}
\providecommand{\urlprefix}{URL }
\expandafter\ifx\csname urlstyle\endcsname\relax
  \providecommand{\doi}[1]{doi:\discretionary{}{}{}#1}\else
  \providecommand{\doi}{doi:\discretionary{}{}{}\begingroup
  \urlstyle{rm}\Url}\fi
\providecommand{\eprint}[2][]{\url{#2}}

\bibitem{Riess_1998}
A.~G. Riess, A.~V. Filippenko, P.~Challis, A.~Clocchiatti, A.~Diercks, P.~M.
  Garnavich, R.~L. Gilliland, C.~J. Hogan, S.~Jha, R.~P. Kirshner,
  B.~Leibundgut, M.~M. Phillips \emph{et~al.},
\newblock \emph{Observational evidence from supernovae for an accelerating
  universe and a cosmological constant},
\newblock The Astronomical Journal \textbf{116}(3), 1009–1038 (1998),
\newblock \doi{10.1086/300499}.

\bibitem{Perlmutter_1999}
S.~Perlmutter, G.~Aldering, G.~Goldhaber, R.~A. Knop, P.~Nugent, P.~G. Castro,
  S.~Deustua, S.~Fabbro, A.~Goobar, D.~E. Groom, I.~M. Hook, A.~G. Kim
  \emph{et~al.},
\newblock \emph{{Measurements of $\Omega$ and $\Lambda$ from 42 High-Redshift
  Supernovae}},
\newblock The Astrophysical Journal \textbf{517}(2), 565–586 (1999),
\newblock \doi{10.1086/307221}.

\bibitem{Carleo:2019ptp}
G.~Carleo, I.~Cirac, K.~Cranmer, L.~Daudet, M.~Schuld, N.~Tishby,
  L.~Vogt-Maranto and L.~Zdeborov{\'a},
\newblock \emph{{Machine learning and the physical sciences}},
\newblock Rev. Mod. Phys. \textbf{91}(4), 045002 (2019),
\newblock \doi{10.1103/RevModPhys.91.045002},
\newblock \eprint{1903.10563}.

\bibitem{LIGOScientific:2019hgc}
{LIGO Scientific and Virgo Collaborations},
\newblock \emph{{A guide to LIGO{\textendash}Virgo detector noise and
  extraction of transient gravitational-wave signals}},
\newblock Class. Quant. Grav. \textbf{37}(5), 055002 (2020),
\newblock \doi{10.1088/1361-6382/ab685e},
\newblock \eprint{1908.11170}.

\bibitem{CMS:2020ebo}
{CMS Collaboration},
\newblock \emph{{Pileup mitigation at CMS in 13 TeV data}},
\newblock JINST \textbf{15}(09), P09018 (2020),
\newblock \doi{10.1088/1748-0221/15/09/P09018},
\newblock \eprint{2003.00503}.

\bibitem{LEPS:2003wug}
{LEPS Collaboration},
\newblock \emph{{Evidence for a narrow S = +1 baryon resonance in
  photoproduction from the neutron}},
\newblock Phys. Rev. Lett. \textbf{91}, 012002 (2003),
\newblock \doi{10.1103/PhysRevLett.91.012002},
\newblock \eprint{hep-ex/0301020}.

\bibitem{ATLAS:2016gzy}
{ATLAS Collaboration},
\newblock \emph{{Search for resonances in diphoton events at $\sqrt{s}$=13 TeV
  with the ATLAS detector}},
\newblock JHEP \textbf{09}, 001 (2016),
\newblock \doi{10.1007/JHEP09(2016)001},
\newblock \eprint{1606.03833}.

\bibitem{Lyons:2013yja}
L.~Lyons,
\newblock \emph{{Discovering the Significance of 5 sigma}}  (2013),
\newblock \eprint{1310.1284}.

\bibitem{Planck:2014dmk}
{Planck Collaboration},
\newblock \emph{{Planck intermediate results. XXX. The angular power spectrum
  of polarized dust emission at intermediate and high Galactic latitudes}},
\newblock Astron. Astrophys. \textbf{586}, A133 (2016),
\newblock \doi{10.1051/0004-6361/201425034},
\newblock \eprint{1409.5738}.

\bibitem{opera_neutrinos_2012}
{OPERA Collaboration},
\newblock \emph{Measurement of the neutrino velocity with the {OPERA} detector
  in the {CNGS} beam},
\newblock Journal of High Energy Physics \textbf{2012}(10) (2012),
\newblock \doi{10.1007/jhep10(2012)093}.

\bibitem{osti_826696}
J.~Friedman,
\newblock \emph{On multivariate goodness-of-fit and two-sample testing}
  (2004),
\newblock \doi{10.2172/826696}.

\bibitem{lopezpaz2018revisitingclassifiertwosampletests}
D.~Lopez-Paz and M.~Oquab,
\newblock \emph{Revisiting classifier two-sample tests} (2018),
  \eprint{1610.06545}.

\bibitem{bonneel:hal-00881872}
N.~Bonneel, J.~Rabin, G.~Peyr{\'e} and H.~Pfister,
\newblock \emph{{Sliced and Radon Wasserstein Barycenters of Measures}},
\newblock {Journal of Mathematical Imaging and Vision} \textbf{51}(1), 22
  (2015),
\newblock \doi{10.1007/s10851-014-0506-3}.

\bibitem{JMLR:v13:gretton12a}
A.~Gretton, K.~M. Borgwardt, M.~J. Rasch, B.~Sch{{\"o}}lkopf and A.~Smola,
\newblock \emph{A kernel two-sample test},
\newblock Journal of Machine Learning Research \textbf{13}(25), 723 (2012).

\bibitem{Kansal:2022spb}
R.~Kansal, A.~Li, J.~Duarte, N.~Chernyavskaya, M.~Pierini, B.~Orzari and
  T.~Tomei,
\newblock \emph{{Evaluating generative models in high energy physics}},
\newblock Phys. Rev. D \textbf{107}(7), 076017 (2023),
\newblock \doi{10.1103/PhysRevD.107.076017},
\newblock \eprint{2211.10295}.

\bibitem{Verde:2023lmm}
L.~Verde, N.~Sch{\"o}neberg and H.~Gil-Mar{\'\i}n,
\newblock \emph{{A Tale of Many H0}},
\newblock Ann. Rev. Astron. Astrophys. \textbf{62}(1), 287 (2024),
\newblock \doi{10.1146/annurev-astro-052622-033813},
\newblock \eprint{2311.13305}.

\bibitem{DelDebbio:2021whr}
L.~Del~Debbio, T.~Giani and M.~Wilson,
\newblock \emph{{Bayesian approach to inverse problems: an application to NNPDF
  closure testing}},
\newblock Eur. Phys. J. C \textbf{82}(4), 330 (2022),
\newblock \doi{10.1140/epjc/s10052-022-10297-x},
\newblock \eprint{2111.05787}.

\bibitem{AnauMontel:2024flo}
N.~Anau~Montel, J.~Alvey and C.~Weniger,
\newblock \emph{{Tests for model misspecification in simulation-based
  inference: From local distortions to global model checks}},
\newblock Phys. Rev. D \textbf{111}(8), 083013 (2025),
\newblock \doi{10.1103/PhysRevD.111.083013},
\newblock \eprint{2412.15100}.

\bibitem{Barontini:2025lnl}
A.~Barontini, M.~N. Costantini, G.~De~Crescenzo, S.~Forte and M.~Ubiali,
\newblock \emph{{Evaluating the faithfulness of PDF uncertainties in the
  presence of inconsistent data}}  (2025),
\newblock \eprint{2503.17447}.

\bibitem{Barlow:2002yb}
R.~Barlow,
\newblock \emph{{Systematic errors: Facts and fictions}},
\newblock In \emph{{Conference on Advanced Statistical Techniques in Particle
  Physics}}, pp. 134--144 (2002), \eprint{hep-ex/0207026}.

\bibitem{annurev.nucl.55.090704.151521}
J.~R. Klein and A.~Roodman,
\newblock \emph{{Blind analysis in nuclear and particle physics}},
\newblock Ann. Rev. Nucl. Part. Sci. \textbf{55}, 141 (2005),
\newblock \doi{10.1146/annurev.nucl.55.090704.151521}.

\bibitem{Kuchibhotla:2022}
A.~K. Kuchibhotla, J.~E. Kolassa and T.~A. Kuffner,
\newblock \emph{Post-selection inference},
\newblock Annual Review of Statistics and Its Application \textbf{9}, 505
  (2022),
\newblock \doi{10.1146/annurev-statistics-100421-044639}.

\bibitem{cranmer2015practicalstatisticslhc}
K.~Cranmer,
\newblock \emph{{Practical Statistics for the LHC}} (2015),
  \eprint{1503.07622}.

\bibitem{DeCastro:2018psv}
P.~De~Castro and T.~Dorigo,
\newblock \emph{{INFERNO: Inference-Aware Neural Optimisation}},
\newblock Comput. Phys. Commun. \textbf{244}, 170 (2019),
\newblock \doi{10.1016/j.cpc.2019.06.007},
\newblock \eprint{1806.04743}.

\bibitem{Simpson:2022suz}
N.~Simpson and L.~Heinrich,
\newblock \emph{{neos: End-to-End-Optimised Summary Statistics for High Energy
  Physics}},
\newblock J. Phys. Conf. Ser. \textbf{2438}(1), 012105 (2023),
\newblock \doi{10.1088/1742-6596/2438/1/012105},
\newblock \eprint{2203.05570}.

\bibitem{doi:10.1073/pnas.1912789117}
K.~Cranmer, J.~Brehmer and G.~Louppe,
\newblock \emph{The frontier of simulation-based inference},
\newblock Proceedings of the National Academy of Sciences \textbf{117}(48),
  30055 (2020),
\newblock \doi{10.1073/pnas.1912789117}.

\bibitem{Cannon:2022}
P.~Cannon, D.~Ward and S.~M. Schmon,
\newblock \emph{Investigating the impact of model misspecification in neural
  simulation-based inference} (2022), \eprint{2209.01845}.

\bibitem{Wehenkel:2024}
A.~Wehenkel, J.~L. Gamella, O.~Sener, J.~Behrmann, G.~Sapiro, J.-H. Jacobsen
  and M.~Cuturi,
\newblock \emph{Addressing misspecification in simulation-based inference
  through data-driven calibration},
\newblock In \emph{Proceedings of the 42nd International Conference on Machine
  Learning}, ICML'25. JMLR.org (2025), \eprint{2405.08719}.

\bibitem{Pierre:2025ulp}
S.~Pierre, B.~R.-S. Blancard, C.~Hahn and M.~Eickenberg,
\newblock \emph{{Mitigating model misspecification in simulation-based
  inference for galaxy clustering}},
\newblock Phys. Rev. D \textbf{113}(4), 043536 (2026),
\newblock \doi{10.1103/gypc-sqnx},
\newblock \eprint{2507.03086}.

\bibitem{Dax:2026qsm}
M.~Dax, T.~Heimel and G.~Louppe,
\newblock \emph{{An Introduction to Bayesian and Frequentist Simulation-Based
  Inference with Machine Learning}}  (2026),
\newblock \eprint{2607.21702}.

\bibitem{Dorigo:2020ldg}
T.~Dorigo and P.~De~Castro~Manzano,
\newblock \emph{{Dealing with Nuisance Parameters using Machine Learning in
  High Energy Physics: a Review}}  (2020),
\newblock \eprint{2007.09121}.

\bibitem{Haussmann:2026gbi}
M.~Hau{\ss}mann, R.~Winterhalder and M.~Ubiali,
\newblock \emph{{Uncertainty in Physics and AI: Taxonomy, Quantification, and
  Validation}}  (2026),
\newblock \eprint{2605.10378}.

\bibitem{Sinervo:2003}
P.~K. Sinervo,
\newblock \emph{Definition and treatment of systematic uncertainties in high
  energy physics and astrophysics},
\newblock In \emph{Proceedings of PHYSTAT2003 (SLAC)} (2003),
  \eprint{2510.24313}.

\bibitem{Bagnaschi:2014wea}
E.~Bagnaschi, M.~Cacciari, A.~Guffanti and L.~Jenniches,
\newblock \emph{{An extensive survey of the estimation of uncertainties from
  missing higher orders in perturbative calculations}},
\newblock JHEP \textbf{02}, 133 (2015),
\newblock \doi{10.1007/JHEP02(2015)133},
\newblock \eprint{1409.5036}.

\bibitem{Duhr:2021mfd}
C.~Duhr, A.~Huss, A.~Mazeliauskas and R.~Szafron,
\newblock \emph{{An analysis of Bayesian estimates for missing higher orders in
  perturbative calculations}},
\newblock JHEP \textbf{09}, 122 (2021),
\newblock \doi{10.1007/JHEP09(2021)122},
\newblock \eprint{2106.04585}.

\bibitem{Tackmann:2024kci}
F.~J. Tackmann,
\newblock \emph{{Beyond scale variations: perturbative theory uncertainties
  from nuisance parameters}},
\newblock JHEP \textbf{08}, 098 (2025),
\newblock \doi{10.1007/JHEP08(2025)098},
\newblock \eprint{2411.18606}.

\bibitem{ATL-PHYS-PUB-2019-043}
{ATLAS Collaboration},
\newblock \emph{{Study of $t\bar{t}b\bar{b}$ and $t\bar{t}W$ background
  modelling for $t\bar{t}H$ analyses}}  (2019),
\newblock \href{https://cds.cern.ch/record/2697143}{ATL-PHYS-PUB-2019-043}.

\bibitem{cranmer2016approximatinglikelihoodratioscalibrated}
K.~Cranmer, J.~Pavez and G.~Louppe,
\newblock \emph{Approximating likelihood ratios with calibrated discriminative
  classifiers} (2016), \eprint{1506.02169}.

\bibitem{Rogozhnikov_2016}
A.~Rogozhnikov,
\newblock \emph{Reweighting with boosted decision trees},
\newblock Journal of Physics: Conference Series \textbf{762}, 012036 (2016),
\newblock \doi{10.1088/1742-6596/762/1/012036}.

\bibitem{algren2023flowawaydifferencesconditional}
M.~Algren, T.~Golling, M.~Guth, C.~Pollard and J.~A. Raine,
\newblock \emph{Flow away your differences: Conditional normalizing flows as an
  improvement to reweighting} (2023), \eprint{2304.14963}.

\bibitem{Daumann_2024}
C.~Daumann, M.~Donega, J.~Erdmann, M.~Galli, J.~L. Späh and D.~Valsecchi,
\newblock \emph{One flow to correct them all: Improving simulations in
  high-energy physics with a single normalising flow and a switch},
\newblock Computing and Software for Big Science \textbf{8}(1) (2024),
\newblock \doi{10.1007/s41781-024-00125-0}.

\bibitem{Pollard:2021fqv}
C.~Pollard and P.~Windischhofer,
\newblock \emph{{Transport away your problems: Calibrating stochastic
  simulations with optimal transport}},
\newblock Nucl. Instrum. Meth. A \textbf{1027}, 166119 (2022),
\newblock \doi{10.1016/j.nima.2021.166119},
\newblock \eprint{2107.08648}.

\bibitem{Algren:2025zff}
M.~Algren, T.~Golling, F.~A. Di~Bello and C.~Pollard,
\newblock \emph{{Mind the Gap: Navigating Inference with Optimal Transport
  Maps}}  (2025),
\newblock \eprint{2507.08867}.

\bibitem{ATLAS:2025rbr}
{ATLAS Collaboration},
\newblock \emph{{A continuous calibration of the ATLAS flavour-tagging
  classifiers via optimal transportation maps}},
\newblock Eur. Phys. J. C \textbf{85}(11), 1272 (2025),
\newblock \doi{10.1140/epjc/s10052-025-14682-0},
\newblock \eprint{2505.13063}.

\bibitem{Assi:2025avy}
B.~Assi, C.~Bierlich, P.~Ilten, T.~Menzo, S.~Mrenna, M.~Szewc, M.~K. Wilkinson,
  A.~Youssef and J.~Zupan,
\newblock \emph{{Characterizing the hadronization of parton showers using the
  HOMER method}},
\newblock SciPost Phys. \textbf{19}(5), 125 (2025),
\newblock \doi{10.21468/SciPostPhys.19.5.125},
\newblock \eprint{2503.05667}.

\bibitem{Huetsch:2024quz}
N.~Huetsch, J.~{Mariño Villadamigo}, A.~Shmakov, S.~Diefenbacher, V.~Mikuni,
  T.~Heimel, M.~Fenton, K.~Greif, B.~Nachman, D.~Whiteson, A.~Butter and
  T.~Plehn,
\newblock \emph{{The landscape of unfolding with machine learning}},
\newblock SciPost Phys. \textbf{18}(2), 070 (2025),
\newblock \doi{10.21468/SciPostPhys.18.2.070},
\newblock \eprint{2404.18807}.

\bibitem{Backes:2022sph}
M.~Backes, A.~Butter, M.~Dunford and B.~Malaescu,
\newblock \emph{{An unfolding method based on conditional invertible neural
  networks (cINN) using iterative training}},
\newblock SciPost Phys. Core \textbf{7}(1), 007 (2024),
\newblock \doi{10.21468/scipostphyscore.7.1.007},
\newblock \eprint{2212.08674}.

\bibitem{Butter:2025via}
A.~Butter, T.~Heimel, N.~Huetsch, M.~Kagan and T.~Plehn,
\newblock \emph{{Simulation-Prior Independent Neural Unfolding Procedure}}
  (2025),
\newblock \eprint{2507.15084}.

\bibitem{Louppe:2016ylz}
G.~Louppe, M.~Kagan and K.~Cranmer,
\newblock \emph{{Learning to Pivot with Adversarial Networks}},
\newblock In \emph{{Advances in Neural Information Processing Systems 30}}
  (2017), \eprint{1611.01046}.

\bibitem{Kasieczka:2020yyl}
G.~Kasieczka and D.~Shih,
\newblock \emph{{Robust Jet Classifiers through Distance Correlation}},
\newblock Phys. Rev. Lett. \textbf{125}(12), 122001 (2020),
\newblock \doi{10.1103/PhysRevLett.125.122001},
\newblock \eprint{2001.05310}.

\bibitem{Ghosh:2021roe}
A.~Ghosh, B.~Nachman and D.~Whiteson,
\newblock \emph{{Uncertainty-aware machine learning for high energy physics}},
\newblock Phys. Rev. D \textbf{104}(5), 056026 (2021),
\newblock \doi{10.1103/PhysRevD.104.056026},
\newblock \eprint{2105.08742}.

\bibitem{Baldi:2016fzo}
P.~Baldi, K.~Cranmer, T.~Faucett, P.~Sadowski and D.~Whiteson,
\newblock \emph{{Parameterized neural networks for high-energy physics}},
\newblock Eur. Phys. J. C \textbf{76}(5), 235 (2016),
\newblock \doi{10.1140/epjc/s10052-016-4099-4},
\newblock \eprint{1601.07913}.

\bibitem{Wilkinson:2025nxv}
A.~Wilkinson, R.~Radev and S.~Alonso-Monsalve,
\newblock \emph{{Contrastive learning for robust representations of neutrino
  data}},
\newblock Phys. Rev. D \textbf{111}(9), 092011 (2025),
\newblock \doi{10.1103/PhysRevD.111.092011},
\newblock \eprint{2502.07724}.

\bibitem{Dillon:2021gag}
B.~M. Dillon, G.~Kasieczka, H.~Olischlager, T.~Plehn, P.~Sorrenson and
  L.~Vogel,
\newblock \emph{{Symmetries, safety, and self-supervision}},
\newblock SciPost Phys. \textbf{12}(6), 188 (2022),
\newblock \doi{10.21468/SciPostPhys.12.6.188},
\newblock \eprint{2108.04253}.

\bibitem{Harris:2024sra}
P.~Harris, J.~Krupa, M.~Kagan, B.~Maier and N.~Woodward,
\newblock \emph{{Resimulation-based self-supervised learning for pretraining
  physics foundation models}},
\newblock Phys. Rev. D \textbf{111}(3), 032010 (2025),
\newblock \doi{10.1103/PhysRevD.111.032010},
\newblock \eprint{2403.07066}.

\bibitem{Golling:2024abg}
T.~Golling, L.~Heinrich, M.~Kagan, S.~Klein, M.~Leigh, M.~Osadchy and J.~A.
  Raine,
\newblock \emph{{Masked particle modeling on sets: towards self-supervised high
  energy physics foundation models}},
\newblock Mach. Learn. Sci. Tech. \textbf{5}(3), 035074 (2024),
\newblock \doi{10.1088/2632-2153/ad64a8},
\newblock \eprint{2401.13537}.

\bibitem{VERaiPHY_representation}
S.~Klein and D.~Muthukrishna,
\newblock \emph{{Representation Learning in Fundamental Physics}},
\newblock In preparation.

\bibitem{Fenton:2020woz}
M.~J. Fenton, A.~Shmakov, T.-W. Ho, S.-C. Hsu, D.~Whiteson and P.~Baldi,
\newblock \emph{{Permutationless many-jet event reconstruction with symmetry
  preserving attention networks}},
\newblock Phys. Rev. D \textbf{105}(11), 112008 (2022),
\newblock \doi{10.1103/PhysRevD.105.112008},
\newblock \eprint{2010.09206}.

\bibitem{pmlr-v119-bogatskiy20a}
A.~Bogatskiy, B.~Anderson, J.~Offermann, M.~Roussi, D.~Miller and R.~Kondor,
\newblock \emph{{L}orentz group equivariant neural network for particle
  physics},
\newblock In \emph{Proceedings of the 37th International Conference on Machine
  Learning}, vol. 119 of \emph{Proceedings of Machine Learning Research}, pp.
  992--1002. PMLR (2020).

\bibitem{Gong:2022lye}
S.~Gong, Q.~Meng, J.~Zhang, H.~Qu, C.~Li, S.~Qian, W.~Du, Z.-M. Ma and T.-Y.
  Liu,
\newblock \emph{{An efficient Lorentz equivariant graph neural network for jet
  tagging}},
\newblock JHEP \textbf{07}, 030 (2022),
\newblock \doi{10.1007/JHEP07(2022)030},
\newblock \eprint{2201.08187}.

\bibitem{Brehmer:2024yqw}
J.~Brehmer, V.~Bres{\'o}, P.~de~Haan, T.~Plehn, H.~Qu, J.~Spinner and
  J.~Thaler,
\newblock \emph{{A Lorentz-equivariant transformer for all of the LHC}},
\newblock SciPost Phys. \textbf{19}(4), 108 (2025),
\newblock \doi{10.21468/SciPostPhys.19.4.108},
\newblock \eprint{2411.00446}.

\bibitem{Bogatskiy:2022czk}
A.~Bogatskiy, T.~Hoffman, D.~W. Miller and J.~T. Offermann,
\newblock \emph{{PELICAN: Permutation Equivariant and Lorentz Invariant or
  Covariant Aggregator Network for Particle Physics}}  (2022),
\newblock \eprint{2211.00454}.

\bibitem{hepmllivingreview}
{HEP ML Community},
\newblock \emph{{A Living Review of Machine Learning for Particle Physics}},
\newblock
  \href{https://iml-wg.github.io/HEPML-LivingReview/}{https://iml-wg.github.io/HEPML-LivingReview/}.

\bibitem{VERaiPHY_symmetries}
J.~Spinner and S.~Villar,
\newblock \emph{{Symmetry-Informed Machine Learning for Fundamental Physics}},
\newblock In preparation.

\bibitem{ATLAS:2011kws}
{ATLAS Collaboration},
\newblock \emph{{A measurement of the ratio of the $W$ and $Z$ cross sections
  with exactly one associated jet in $pp$ collisions at $\sqrt{s}=$7 TeV with
  ATLAS}},
\newblock Phys. Lett. B \textbf{708}, 221 (2012),
\newblock \doi{10.1016/j.physletb.2012.01.042},
\newblock \eprint{1108.4908}.

\bibitem{ATLAS:2017ayi}
{ATLAS Collaboration},
\newblock \emph{{Search for new phenomena in high-mass diphoton final states
  using 37 fb$^{-1}$ of proton--proton collisions collected at $\sqrt{s}=13$
  TeV with the ATLAS detector}},
\newblock Phys. Lett. B \textbf{775}, 105 (2017),
\newblock \doi{10.1016/j.physletb.2017.10.039},
\newblock \eprint{1707.04147}.

\bibitem{CMS:2016kgr}
{CMS Collaboration},
\newblock \emph{{Search for high-mass diphoton resonances in
  proton{\textendash}proton collisions at 13 TeV and combination with 8 TeV
  search}},
\newblock Phys. Lett. B \textbf{767}, 147 (2017),
\newblock \doi{10.1016/j.physletb.2017.01.027},
\newblock \eprint{1609.02507}.

\bibitem{Frate:2017mai}
M.~Frate, K.~Cranmer, S.~Kalia, A.~Vandenberg-Rodes and D.~Whiteson,
\newblock \emph{{Modeling Smooth Backgrounds and Generic Localized Signals with
  Gaussian Processes}}  (2017),
\newblock \eprint{1709.05681}.

\bibitem{Hallin:2021wme}
A.~Hallin, J.~Isaacson, G.~Kasieczka, C.~Krause, B.~Nachman, T.~Quadfasel,
  M.~Schlaffer, D.~Shih and M.~Sommerhalder,
\newblock \emph{{Classifying anomalies through outer density estimation}},
\newblock Phys. Rev. D \textbf{106}(5), 055006 (2022),
\newblock \doi{10.1103/PhysRevD.106.055006},
\newblock \eprint{2109.00546}.

\bibitem{Metodiev:2017vrx}
E.~M. Metodiev, B.~Nachman and J.~Thaler,
\newblock \emph{{Classification without labels: Learning from mixed samples in
  high energy physics}},
\newblock JHEP \textbf{10}, 174 (2017),
\newblock \doi{10.1007/JHEP10(2017)174},
\newblock \eprint{1708.02949}.

\bibitem{Kasieczka:2020pil}
G.~Kasieczka, B.~Nachman, M.~D. Schwartz and D.~Shih,
\newblock \emph{{Automating the ABCD method with machine learning}},
\newblock Phys. Rev. D \textbf{103}(3), 035021 (2021),
\newblock \doi{10.1103/PhysRevD.103.035021},
\newblock \eprint{2007.14400}.

\bibitem{ATLAS:2022swp}
{ATLAS Collaboration},
\newblock \emph{{Tools for estimating fake/non-prompt lepton backgrounds with
  the ATLAS detector at the LHC}},
\newblock JINST \textbf{18}(11), T11004 (2023),
\newblock \doi{10.1088/1748-0221/18/11/T11004},
\newblock \eprint{2211.16178}.

\bibitem{novikov2025alphaevolvecodingagentscientific}
A.~Novikov, N.~Vũ, M.~Eisenberger, E.~Dupont, P.-S. Huang, A.~Z. Wagner,
  S.~Shirobokov, B.~Kozlovskii, F.~J.~R. Ruiz, A.~Mehrabian, M.~P. Kumar,
  A.~See \emph{et~al.},
\newblock \emph{{AlphaEvolve}: A coding agent for scientific and algorithmic
  discovery} (2025), \eprint{2506.13131}.

\bibitem{wahl2026probabilisticframeworkllmbasedmodel}
S.~Wahl, R.~Schenk, A.~Farnoud, J.~H. Macke and D.~Gedon,
\newblock \emph{{A Probabilistic Framework for LLM-Based Model Discovery}}
  (2026), \eprint{2602.18266}.

\bibitem{georgiev2025mathematicalexplorationdiscoveryscale}
B.~Georgiev, J.~Gómez-Serrano, T.~Tao and A.~Z. Wagner,
\newblock \emph{Mathematical exploration and discovery at scale} (2025),
  \eprint{2511.02864}.

\end{thebibliography}
\end{document}